\documentclass{article}

\usepackage{graphicx}
\usepackage{amssymb}
\usepackage{amsmath}
\usepackage{cancel}

\def\ihat{{\bf \hat x}}
\def\jhat{{\bf \hat y}}
\def\khat{{\bf \hat z}}

\def\phat{{\hat{\mbox{\boldmath$\phi$}}}}

\def\rcurs{{\mbox{$\resizebox{.16in}{.08in}{\includegraphics{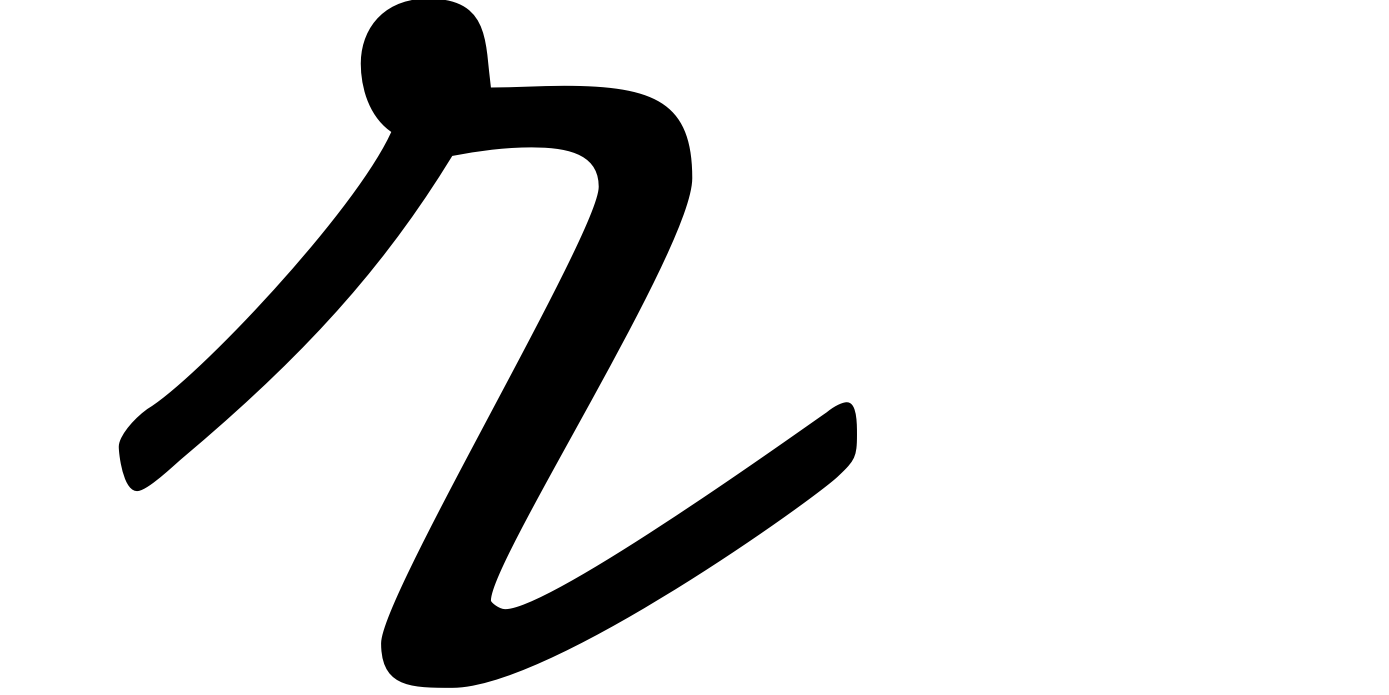}}$}}}
\def\brcurs{{\mbox{$\resizebox{.16in}{.08in}{\includegraphics{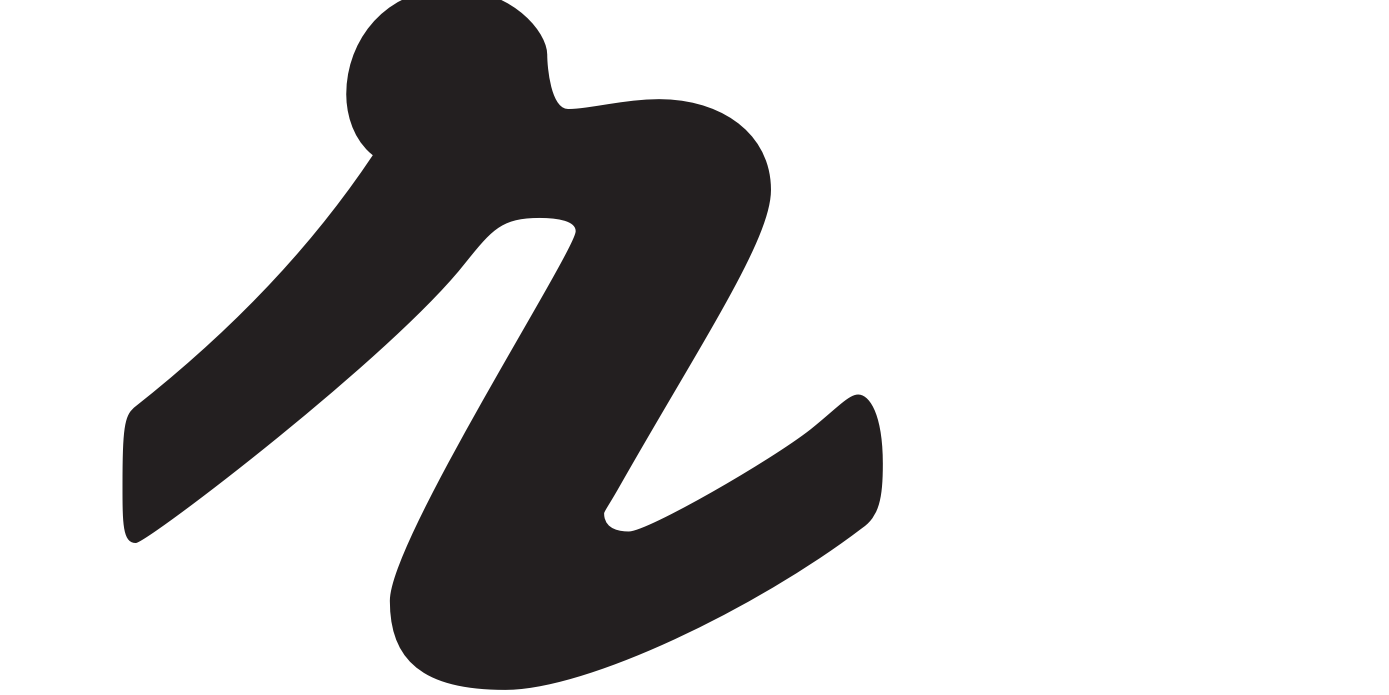}}$}}}
\def\hrcurs{{\mbox{$\hat \brcurs$}}}

\begin{document}

\title{The Fields of a Charged Particle in Hyperbolic Motion}

\author{Joel Franklin\\
David J.~Griffiths\\
{\it Department of Physics,
Reed College}\\{\it  Portland, Oregon  97202}}

\maketitle

\begin{abstract}  
A particle in hyperbolic motion produces electric fields that appear to terminate in mid-air, violating Gauss's law.   The resolution to this paradox has been known for sixty years, but exactly why the naive approach fails is not so clear.
\end{abstract}

\section{Introduction}  

In special relativity a particle of mass $m$ subject to a constant force $F$ undergoes ``hyperbolic motion":
\begin{equation}
z(t) = \sqrt{b^2+(ct)^2},
\end{equation}
where $b \equiv mc^2/F$.  The particle flies in from infinity along (say) the $z$ axis, comes to rest at $z(0)= b$, and returns to infinity; its velocity approaches $\pm c$ asymptotically as $t\to \pm \infty$ (Figure 1).  

\vskip0in
\begin{figure}[t]
\hskip.4in\scalebox{.4}[.475]{\includegraphics{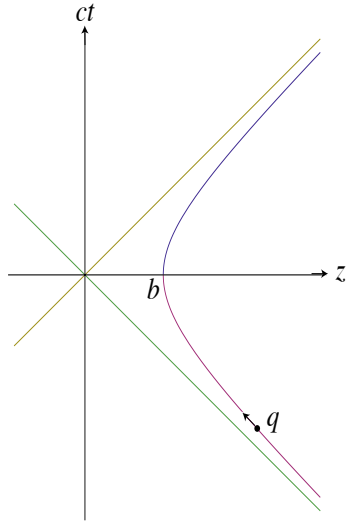}}
\vskip-1in
\caption{Hyperbolic motion.}
\end{figure}

Because information cannot travel faster than the speed of light, the region below the main diagonal ($z=-ct$) is ignorant of the particle's existence---the particle is ``over the horizon."  For someone at the origin it first comes into view at $t=0$.  If the particle is electrically charged, its fields are necessarily zero for all $z<0$, at time $t=0$.  But the electric field for $z>0$ is {\it not} zero, and as we shall see the field lines appear to terminate in mid-air at the $xy$ plane.\footnote{The {\it entire} $xy$ plane first ``sees" the charge at time $t=0$.  If this seems surprising, refer to Appendix A.}  This would violate Gauss's law; it cannot be true.  Our task is to locate the error and fix it.\footnote{The fields of a point charge in hyperbolic motion were first considered by M.~Born, ``Die Theorie des starren Elektrons in der Kinematik des Relativit\"atsprinzips," {\it Ann.~Physik} {\bf 30}, 1-56 (1909).  For the early history of the problem see W.~Pauli, {\it Theory of Relativity} (reprint by Dover, New York, 1981), Section 32($\gamma$).  For a comprehensive history see E.~Eriksen and {\O}.~Gr{\o}n, ``Electrodynamics of Hyperbolically Accelerated Charges.  I.~The Electromagnetic Field of a Charged Particle with Hyperbolic Motion," {\it Ann.~Phys.} {\bf 286}, 320-342 (2000).  See also S.~Lyle, {\it Uniformly Accelerating Charged Particles: A Threat to the Equivalence Principle} (Springer, Berlin, 2008).}

In Section 2 we calculate the electric field of a charge $q$ in hyperbolic motion, at time $t=0$.  A plot of the field lines shows that they do not go continuously to zero at the $xy$ plane.  In Section 3 we explore the case of ``truncated" hyperbolic motion (hyperbolic motion back to time $t=-t_0$, adjoined to constant velocity for earlier times).  In this case the field lines make a sharp turn as they approach the $xy$ plane, and there is {\it no} violation of Gauss's law.  In Section 4 we work out the potentials for a charge in hyperbolic motion, finding once again that we must adjoin ``by hand" a term inspired by the truncated case.  In Section 5 we ask how the naive calculations missed the extra term, and conclude with the puzzle unresolved.  Appendices A and B supply some algebraic details, and Appendix C examines the radiation from a charge in hyperbolic motion; surprisingly, the ``extra" terms do not contribute.  

\section{Electric Field of a Charge in Hyperbolic Motion}

We begin by calculating the electric field at the point ${\bf r}=(x,0,z)$, with $z>0$.  According to the standard formula,\footnote{See, for example, D.~J.~Griffiths, {\it Introduction to Electrodynamics, 4th ed.} (Pearson, Upper Saddle River, NJ, 2013), Eq.~10.72.}
\begin{equation}
{\bf E}({\bf r},t) =\frac{q}{4\pi\epsilon_0}\frac{\rcurs}{(\brcurs \cdot {\bf u})^3}\left[(c^2-v^2){\bf u} + \brcurs\times({\bf u} \times {\bf a})\right],
\end{equation}
where 
\begin{equation}
\brcurs = x\,\ihat + \left(z\ -\sqrt{b^2+(ct_r)^2}\right)\,\khat,
\end{equation}
\begin{equation}
{\bf u} = c\hrcurs - {\bf v} = \frac{1}{\rcurs}(c\brcurs -\rcurs \hskip-.05in{\bf v}),
\end{equation}
\begin{equation}
{\bf v} = \frac{c^2 t_r}{\sqrt{b^2+(ct_r)^2}}\,\khat,
\end{equation}
and 
\begin{equation}
{\bf a} = \frac{(bc)^2}{\left(\sqrt{b^2+(ct_r)^2}\right)^3}\,\khat.
\end{equation} 
The retarded time, $t_r$, is defined in general by
\begin{equation}
\rcurs = c(t-t_r),
\end{equation}
but for the moment we'll assume $t=0$ (so 
$t_r$ is negative).  Then
\begin{equation}
(ct_r)^2 = x^2 + \left(z-\sqrt{b^2+(ct_r)^2}\right)^2 = x^2+z^2 -2z\sqrt{b^2+(ct_r)^2} + b^2+(ct_r)^2,
\end{equation}
and hence
\begin{equation}
ct_r = -\frac{1}{2z}\sqrt{ \left(x^2+z^2+b^2\right)^2-(2zb)^2}.
\end{equation}

Putting all this together, and simplifying,
\begin{equation}
{\bf E}(x,0,z) = \frac{qb^2}{\pi\epsilon_0}\frac{\left(z^2-x^2-b^2\right)\,\khat + (2xz)\,\ihat}{\left(\sqrt{(z^2+x^2+b^2)^2-(2zb)^2}\right)^3}.
\end{equation}
That is for $z>0$, of course; for $z<0$ the field is zero.  In cylindrical coordinates $(s, \phi, z)$, then\footnote{This field was first obtained by G.~A.~Schott, {\it Electromagnetic Radiation,} Cambridge University Press, Cambridge, UK (1912), pp.~63-69.}
\begin{equation}
{\bf E}(s,\phi,z) = \frac{qb^2}{\pi\epsilon_0}\frac{\left(z^2-s^2-b^2\right)\,\khat + 2z\,{\bf s}}{\left(\sqrt{(z^2+s^2+b^2)^2-(2zb)^2}\right)^3} \,\theta(z),
\end{equation}
where $\theta(z)$ is the step function (1 if $z>0$, otherwise 0).  This field is plotted in Figure 2; the field lines are circles, centered on the $s$ axis and passing through the instantaneous position of the charge.

\vskip0in
\begin{figure}[h]
\hskip1in\scalebox{.5}[.5]{\includegraphics{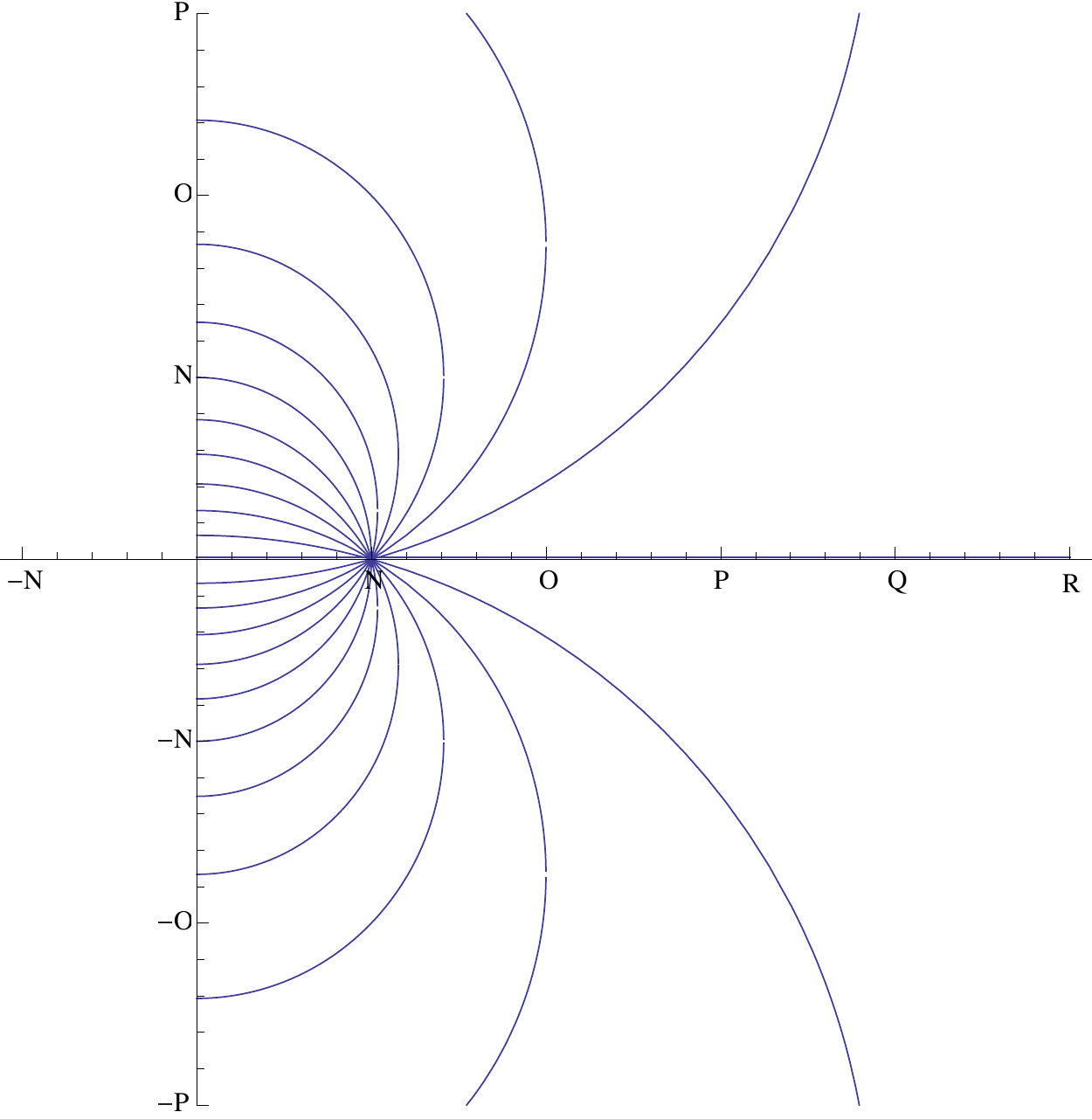}}
\vskip0in
\caption{Field of a particle in hyperbolic motion, with $b=1$ (naive).}
\end{figure}

As required by Gauss's law, $\nabla \cdot {\bf E} =0$ for all $z>0$ (except at the point $s=0, z=b$, where the charge is located).  However, {\bf E} is plainly {\it not} divergenceless at the $xy$ plane, where the field lines terminate in mid-air.  Indeed, the field immediately to the right of the $z=0$ plane is
\begin{equation}
{\bf E}(s,\phi,0^+)  = -\frac{qb^2}{\pi\epsilon_0}\frac{1}{(s^2+b^2)^2}\,\khat,
\end{equation}
and the flux of {\bf E} through a cylindrical Gaussian ``pillbox" of radius $r$, centered at the origin and straddling the plane, with infinitesimal thickness, is
 \begin{equation}
 \int {\bf E} \cdot d{\bf a} = -\frac{qb^2}{\pi\epsilon_0}\int_0^r \frac{1}{(s^2+b^2)^2} 2\pi s\,ds =-\frac{q}{\epsilon_0}\left(\frac{r^2}{r^2+b^2}\right),
 \end{equation}
even though the pillbox encloses no charge.  Something is obviously amiss---we appear to have lost a crucial piece of the field at $z=0$.

\section{Truncated Hyperbolic Motion}
Suppose the acceleration does not extend all the way back to $t=-\infty$, but begins at time $t_0=-\alpha b/c$ (for some $\alpha >0$), when the particle was at
\begin{equation}
z(t_0) = b\sqrt{1+\alpha^2},
\end{equation}
and its velocity was
\begin{equation}
{\bf v}(t_0) = -\frac{\alpha c}{\sqrt{1+\alpha^2}}\,\khat;
\end{equation}
prior to $t_0$ the velocity was constant.  In other words,  replace Eq.~1 with
\begin{equation}
z(t) = \begin{cases}{\displaystyle \frac{1}{\sqrt{1+\alpha^2}}(b-\alpha ct)} & (t<t_0 = -\alpha b/c)\\ &\\\sqrt{b^2+(ct)^2}&(t\geq t_0).\end{cases}
\end{equation}

At time $t=0$, for all points {\it outside} a sphere of radius $r= -ct_0=\alpha b$, centered at $z(t_0)$, the field is that of a charge moving at constant velocity---the ``flattened" Heaviside field\footnote{E.~M.~Purcell and D.~J.~Morin, {\it Electricity and Magnetism, 3rd ed.} (Cambridge University Press, Cambridge, UK, 2013), Section 5.6.} radiating from the place $q$ {\it would} have reached, had it continued on its original flight plan ($b/\sqrt{1+\alpha^2}$):
\begin{equation}
{\bf E} = \frac{q}{4\pi\epsilon_0}\frac{1-(v/c)^2}{[1-(v/c)^2\sin^2\theta]^{3/2}}\frac{{\bf R}}{R^3}.
\end{equation}
The left edge of the sphere is at $\left(\sqrt{1+\alpha^2}-\alpha\right)b$ (which is always positive, but goes to zero as $\alpha\to \infty$).  {\it Inside} the sphere, where news of the acceleration has been received, the field is given by Eq.~11 (Figure 3).  The field lines evidently join up in a thin layer at the surface of the sphere, representing the brief interval during which the motion switches from uniform to hyperbolic.

\vskip-0in
\begin{figure}[h]
\hskip1in\scalebox{.5}[.5]{\includegraphics{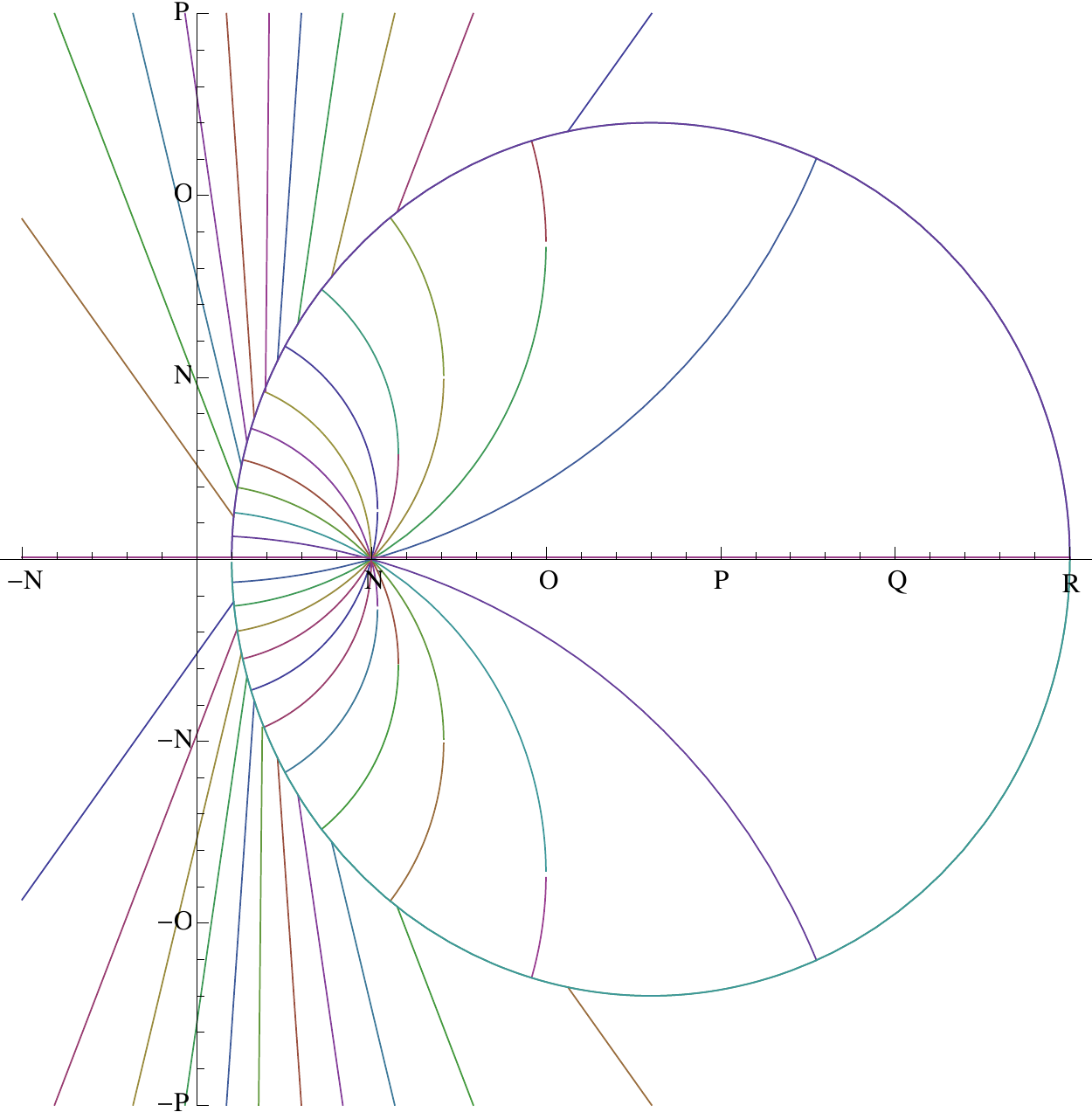}}
\vskip0in
\caption{Field lines for truncated hyperbolic motion ($b=1$, $\alpha = 12/5$).}
\end{figure}

As alpha increases (that is, as $t_0$ recedes into the more distant past), the radius of the sphere increases, and its left surface flattens out against the $xy$ plane.  Meanwhile the ``outside" field compresses into a disk perpendicular to the motion, and  squeezes also onto the $xy$ plane.  The complete field lines now execute a 90$^\circ$ turn at $z=0$, as required to rescue Gauss's law.  Indeed, for $\alpha\to \infty$ the constant velocity portion of the field approaches that of a point charge moving at speed $c$:\footnote{J.~M.~Aguirregabiria, A.~Hern\'andez, and M.~Rivas, ``$\delta$-function converging sequences," {\it Am.~J.~Phys.}~{\bf 70}, 180-185 (2002), Eq.~50.  The fields of a massless point charge are considered also in J.~D.~Jackson, {\it Classical Electrodynamics, 3rd ed.} (Wiley, New York, 1999), Prob.~11.18, W.~Thirring {\it Classical Mathematical Physics: Dynamical Systems and Field Theories} (Springer-Verlag, New York, 1997), pp.~367-368, and M.~V.~Kozyulin and Z.~K.~Silagadze, ``Light bending by a Coulomb field and the Aichelburg-Sexl ultraboost," {\it Eur.~J.~Phys.}~{\bf 32}, 1357-1365 (2011), Eq.~20.}
\begin{equation}
{\bf E}'(s,\phi,z) = \frac{q}{2\pi\epsilon_0}\frac{{\bf s}}{s^2}\delta(z).
\end{equation}
Using the same Gaussian pillbox as before, this field yields
\begin{equation}
\int {\bf E}' \cdot d{\bf a} = \frac{q}{2\pi\epsilon_0}\frac{r}{r^2}(2\pi r)\int\delta(z)\,dz = \frac{q}{\epsilon_0}.
\end{equation}
This is appropriate, of course---had the particle continued at its original velocity ($c$) it would now be inside the box (at the origin).

\vskip0in
\begin{figure}[t]
\hskip1in\scalebox{.6}[.6]{\includegraphics{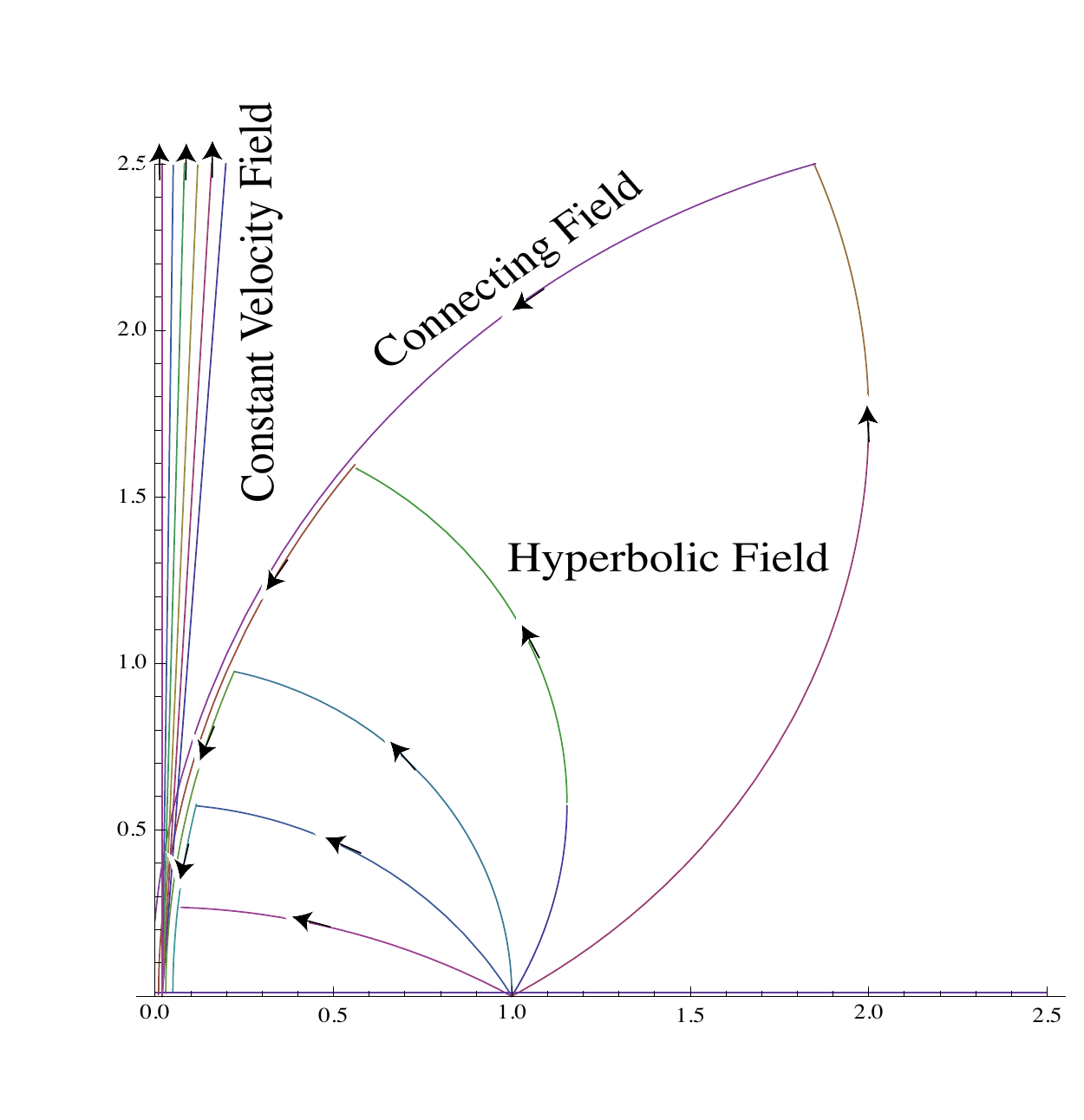}}
\vskip-.2in
\caption{Truncated hyperbolic motion, for large $\alpha$, showing the ``connecting" field.}
\end{figure}

Awkwardly, however, this is {\it not} what was needed to cancel the flux from the hyperbolic part of the field (Eq.~13).  For {\it that} purpose the field on the $xy$ plane should have been 
\begin{equation}
{\bf E}(s,\phi,z) = \frac{q}{2\pi\epsilon_0}\frac{{\bf s}}{s^2+b^2}\delta(z).
\end{equation}
It must be that the ``connecting" field in the spherical shell (the field produced during the transition from uniform to hyperbolic motion), which (in the limit) coincides with the $xy$ plane, and which we have ignored, accounts for the difference, as suggested in Figure 4.  The {\it net} field in the $xy$ plane consists of two parts: the field ${\bf E}'$ due to the portion of the motion at constant velocity, given (in the limit $\alpha\to \infty$) by Eq.~18, and the connecting field that joins it to the hyperbolic part.  It is the {\it sum} of these fields that gives Eq.~20.   The {\it true} field of a charge in hyperbolic motion is evidently\footnote{The delta-function term was first obtained (using a somewhat different method) by H.~Bondi and T.~Gold, ``The field of a uniformly accelerated charge, with special reference to the problem of gravitational acceleration," {\it Proc.~R.~Soc.~London Ser.~A} {\bf 229}, 416-424 (1955).   D.~G.~Boulware, ``Radiation from a Uniformly Accelerated Charge," {\it Ann.~Phys.}\,{\bf 124}, 169-188 (1980) obtained it using the truncated hyperbolic model.  The latter was also explored by W.~Thirring {\it A Course in Mathematical Physics: Classical Field Theory, 2nd ed.} (Springer-Verlag, New York, 1992), p.~78.  See also Lyle, ref.~2, Section 15.9.}
\begin{equation}
{\bf E}(s,\phi,z) =\frac{qb^2}{\pi\epsilon_0}\frac{\left(z^2-s^2-b^2\right)\,\khat + 2z\,{\bf s}}{\left(\sqrt{(z^2+s^2+b^2)^2-(2zb)^2}\right)^3}\theta(z) +  \frac{q}{2\pi\epsilon_0}\frac{{\bf s}}{s^2+b^2}\delta(z)
\end{equation}
and it does not look like Fig.~2, but rather Fig.~5.

\vskip0in
\begin{figure}[t]
\hskip0in\scalebox{.8}[.8]{\includegraphics{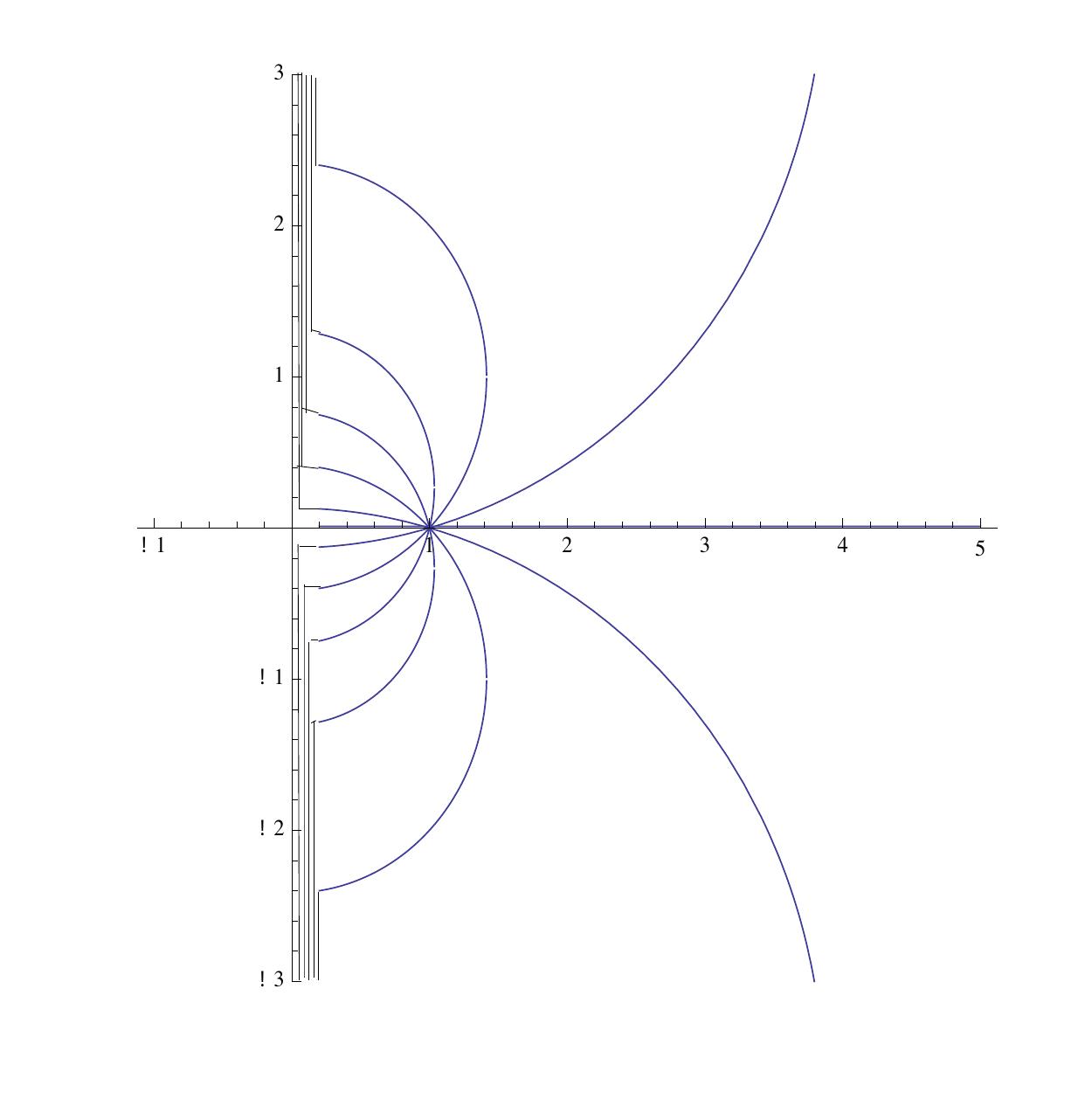}}
\vskip-.3in
\caption{Field of a particle in hyperbolic motion (corrected).}
\end{figure}

As a check, let's calculate the divergence of {\bf E}.  Writing ${\bf E} = {\bf E}_\theta\theta(z) + {\bf E}_\delta$ (in an obvious notation), we have
$$\nabla \cdot [{\bf E}_\theta\theta(z)] = (\nabla\cdot{\bf E}_\theta)\theta(z) + {\bf E}_\theta\cdot [\nabla(\theta)].$$
The first term gives $\rho/\epsilon_0$, for the point charge $q$ at $z=b$; as for the second term,
$$\nabla \theta(z) = \frac{\partial \theta}{\partial z}\,\hat{\bf z} = \delta(z)\,\hat{\bf z},$$
so 
\begin{eqnarray*}
\nabla \cdot \left[{\bf E}_\theta\theta(z)\right]& =&\frac{\rho}{\epsilon_0} + \frac{qb^2}{\pi\epsilon_0}\frac{(z^2-s^2-b^2)}{\left[(z^2+s^2+b^2)^2-(2zb)^2\right]^{3/2}}\,\delta(z) \\
&=& \frac{\rho}{\epsilon_0}-\frac{qb^2}{\pi\epsilon_0}\frac{1}{(s^2+b^2)^2}\,\delta(z).
\end{eqnarray*}
Meanwhile
$$\nabla \cdot {\bf E}_\delta = \frac{q}{2\pi\epsilon_0}\nabla\cdot\left[\frac{{\bf s}}{s^2+b^2}\,\delta(z)\right]=\frac{q}{2\pi\epsilon_0}\frac{1}{s}\frac{\partial}{\partial s}\left[\frac{s^2}{s^2+b^2}\,\delta(z)\right],$$
so
$$\nabla \cdot {\bf E}_\delta = \frac{q}{\pi\epsilon_0}\frac{b^2}{(s^2+b^2)^2}\,\delta(z).$$
This is just right to cancel the extra term in $\nabla \cdot \left[{\bf E}_\theta\theta(z)\right]$, and Gauss's law is sustained:
$$\nabla\cdot{\bf E}=\frac{\rho}{\epsilon_0}.$$

\section{Potential Formulation}
\subsection{Li\'enard-Wiechert Potentials}
The truncated hyperbolic problem guided us to the ``extra" (delta-function) term in Eq.~21, but it does not explain how we missed that term in the first place.  Did it perhaps get lost in going from the potentials to the fields?  Let's work out the Li\'enard-Wiechert potentials,\footnote{Reference 3, Eqs.~10.46 and 10.47.} and calculate the field more carefully:
\begin{equation}
V({\bf r},t) = \frac{q}{4\pi\epsilon_0} \frac{1}{[\rcurs -(\brcurs\cdot {\bf v})/c]},\quad {\bf A}({\bf r},t)= \frac{{\bf v}}{c^2}V({\bf r},t),
\end{equation}
where $\brcurs$ and ${\bf v}$ are evaluated at the retarded time, $t_r$.  For the point ${\bf r} = ({\bf s},z)$, 

\begin{eqnarray}
T_r &=&\frac{-1}{2(T^2-z^2)}\Big[T(s^2+z^2+ b^2-T^2) \nonumber\\
&&\quad\quad\quad-\, z\sqrt{4b^2(T^2-z^2)+(s^2+z^2+b^2-T^2)^2}\Big]
\end{eqnarray}

\noindent ($T \equiv ct$ and $T_r \equiv ct_r$).\footnote{See Appendix B for details of these calculations.   Equation 23 reduces to Eq.~9, of course, when $t=0$ and $s=x$.}  This is for $z>-T$; as we approach the horizon ($z\to -T$), the retarded time goes to $-\infty$, and for $z<-T$ there is no solution with $T>T_r$.

 The scalar potential is
\begin{equation}
V = \frac{q}{4\pi\epsilon_0}\frac{1}{(T^2-z^2)}\left[T - \frac{z(s^2+z^2+b^2-T^2)}{\sqrt{4b^2(T^2-z^2)+(s^2+z^2+b^2-T^2)^2}}\right]\theta(T+z),
\end{equation}
and the vector potential is
\begin{equation}
{\bf A} = \frac{q}{4\pi\epsilon_0 c}\frac{1}{(T^2-z^2)}\left[z-\frac{T(s^2+z^2+b^2-T^2)}{\sqrt{4b^2(T^2-z^2)+(s^2+z^2+b^2-T^2)^2}}\right]\theta(T+z)\,\khat.
\end{equation}
The electric field  is
\begin{eqnarray}
{\bf E} &=& -\nabla V - \frac{\partial{\bf A}}{\partial t} \nonumber\\
&=& \frac{qb^2}{\pi\epsilon_0}\left\{\frac{2z\,{\bf s} - (s^2-z^2 + b^2+T^2)\,\khat}{\left[4b^2(T^2-z^2)+(s^2+z^2+b^2-T^2)^2\right]^{3/2}}\right\}\theta(T+z)
\end{eqnarray}
(which reduces to Eq.~21---{\it without} the extra term---when $t=0$).  Notice that the derivatives of the theta function contribute nothing (we use an overbar to denote the potentials shorn of their $\theta$'s):
\begin{eqnarray}
&&-\delta(T+z) (\bar V+c\bar A_z)=- \frac{q}{4\pi\epsilon_0} \delta(T+z) \frac{1}{(T^2-z^2)}\nonumber\\
&&\times\left[(T+z) -\frac{(T+z)(s^2+z^2+b^2-T^2)}{\sqrt{4b^2(T^2-z^2)+(s^2+z^2+b^2-T^2)^2}}\right]\nonumber\\
&&= -\frac{q}{4\pi\epsilon_0}\frac{\delta(T+z)}{(T-z)}\left[1-\frac{(s^2+b^2)}{(s^2+b^2)}\right]=0.
\end{eqnarray}

Evidently there is something wrong with the Li\'enard-Wiechert potentials themselves; they too are missing a critical term.  To fix them, we play the same game as before: truncate the hyperbolic motion.  We might as well go straight to the limit, with the truncation receding to $-\infty$; we need the potentials of a point charge moving at speed $c$.   There are two candidates in the literature\footnote{R.~Jackiw, D.~Kabat, and M.~Ortiz, ``Electromagnetic fields of a massless particle and the eikonal," {\it Phys.~Lett.} {\bf B 277}, 148-152 (1992).} (which differ by a gauge transformation, though both satisfy the Lorenz condition, $\partial V/\partial t = -c^2(\nabla\cdot {\bf A})$):
\begin{equation}
V_I'= 0,\quad {\bf A}_I' = -\frac{q}{2\pi\epsilon_0c}\frac{{\bf s}}{s^2}\theta(z+T),
\end{equation}
\begin{equation}
V_{II}' = -\frac{q}{2\pi\epsilon_0}\ln\left(\frac{s}{b}\right)\delta(z+T),\quad {\bf A}_{II}' = \frac{q}{2\pi \epsilon_0 c}\ln\left(\frac{s}{b}\right)\delta(z+T)\,\khat
\end{equation}
(in the second case $b$ could actually be {\it any} constant with the dimensions of length, but we might as well use a parameter that is already on the table).

We also need the ``connecting" potentials; our experience with the fields (going from Eq.~18 to Eq.~20) suggests the following ansatz
\begin{equation}
V_I = 0,\quad {\bf A}_I = -\frac{q}{2\pi\epsilon_0c}\frac{{\bf s}}{(s^2+b^2)}\theta(z+T),
\end{equation}
\begin{equation}
V_{II} = -\frac{q}{4\pi\epsilon_0}\ln\left(\frac{s^2+b^2}{b^2}\right)\delta(z+T),\quad {\bf A}_{II} = \frac{q}{4\pi \epsilon_0 c}\ln\left(\frac{s^2+b^2}{b^2}\right)\delta(z+T)\,\khat.
\end{equation}
It is easy to check, in either case, that we recover the correct ``extra" term in the field (Eq.~21).  However, we prefer $V_{II}$ and ${\bf A}_{II}$, because they preserve the Lorenz gauge.\footnote{Potentials 24 and 25 satisfy the Lorenz gauge condition, as does 31, but 30 does not.}

The correct potentials for a point charge in hyperbolic motion are thus\footnote{The delta-function terms in the potentials were first obtained by T.~Fulton and F.~Rohrlich, ``Classical radiation from a uniformly accelerated charge," {\it Ann.~Phys.~}{\bf 9}, 499-517 (1960).} 
\begin{eqnarray}
V &=& \frac{q}{4\pi\epsilon_0}\Bigg\{\frac{1}{(T^2-z^2)}\left[T - \frac{z(s^2+z^2+b^2-T^2)}{\sqrt{4b^2(T^2-z^2)+(s^2+z^2+b^2-T^2)^2}}\right]\theta(T+z)\nonumber\\
&&-\ \ln\left(\frac{s^2+b^2}{b^2}\right)\delta(z+T)\Bigg\},\\
{\bf A} &=& \frac{q}{4\pi\epsilon_0 c}\Bigg\{\frac{1}{(T^2-z^2)}\left[z-\frac{T(s^2+z^2+b^2-T^2)}{\sqrt{4b^2(T^2-z^2)+(s^2+z^2+b^2-T^2)^2}}\right]\theta(T+z)\nonumber\\
&&+\ \ln\left(\frac{s^2+b^2}{b^2}\right)\delta(z+T)\Bigg\}\khat.
\end{eqnarray}
How did the standard Li\'enard-Wiechert construction miss the extra (delta function) terms?  Was it perhaps in the derivation of the Li\'enard-Wiechert potentials from the retarded potentials?

\subsection{Retarded Potentials}
Let's take a further step back, then, and examine the retarded potential\footnote{Ref.~3, Eq.~10.19.  Since all we're doing is searching for a missing term, we may as well concentrate on $V$, and set $t=0$.}
\begin{equation}
V(s,z) = \frac{1}{4\pi\epsilon_0}\int\frac{\rho({\bf r}', t_r)}{\rcurs}\,d^3{\bf r}'.
\end{equation}
In this case
\begin{equation}
\rho({\bf r},t) = q\delta^3\left({\bf r} - \sqrt{b^2+(ct)^2}\,\khat\right),
\end{equation}
and we need $\rho({\bf r}', t_r)$, where (for $t=0$)
\begin{equation}
-ct_r=|{\bf r}-{\bf r}'|=\sqrt{(x-x')^2+(y-y')^2+(z-z')^2}.
\end{equation}
Thus
$$\rho({\bf r}',t_r)=q\delta^3\left(x'\,\ihat + y'\,\jhat +z'\,\khat - \sqrt{b^2+(x-x')^2+(y-y')^2+(z-z')^2}\,\khat\right).$$

Because of the delta function, the denominator ($\rcurs=|{\bf r}-{\bf r}'|$) in Eq.~34 comes outside the integral---with ${\bf r}'$, now, at the retarded point (where the argument of the delta-function vanishes).  What remains is
\begin{eqnarray}
Q&\equiv& \int\rho({\bf r}',t_r)\,d^3{\bf r}' \nonumber\\
&=&q\int\delta(x')\delta(y')\delta\left(z'- \sqrt{b^2+(x-x')^2+(y-y')^2+(z-z')^2}\right)\,dx'dy'dz'\nonumber\\
&=&q\int\delta\left(z'- \sqrt{b^2+s^2+(z-z')^2}\right)\,dz' = q\int\delta\left(f(z')\right)\,dz',
\end{eqnarray}
where $s^2= x^2+y^2$, and
\begin{equation}
f(z')\equiv z' -  \sqrt{b^2+s^2+(z-z')^2}.
\end{equation}

The argument of the delta function vanishes when $z'=z_0$, given by $f(z_0)=0$:
$$z_0= \sqrt{b^2+s^2+(z-z_0)^2}, \quad  z_0^2=b^2+s^2+z^2-2zz_0+z_0^2,$$
or
\begin{equation}
z_0 = \frac{1}{2z}\left(s^2+z^2+b^2\right).
\end{equation}
Note that $z_0$ is non-negative, so there is no solution when $z<0$.  Now
$$\frac{df}{dz'} = 1+\frac{(z-z')}{\sqrt{b^2+s^2+(z-z')^2}},$$
so
\begin{equation}
f'(z_0) = 1+\frac{(z-z_0)}{z_0}=\frac{z}{z_0}= \frac{2z^2}{s^2+z^2+b^2},
\end{equation}
and hence
\begin{equation}
\delta\left(f(z')\right) = \frac{1}{|f'(z_0)|}\,\delta(z'-z_0)=\left(\frac{s^2+z^2+b^2}{2z^2}\right)\delta(z'-z_0).
\end{equation}
Thus
\begin{equation}
Q=q\left(\frac{s^2+z^2+b^2}{2z^2}\right)\theta(z).
\end{equation}

The retarded potential is 
\begin{equation}
V=\frac{1}{4\pi\epsilon_0}\frac{Q}{\rcurs},
\end{equation}
and from Eq.~23 (with $t=0$)
\begin{equation}
\rcurs = -ct_r=-T_r=\frac{1}{2z}\sqrt{(s^2+z^2+b^2)^2-(2bz)^2},
\end{equation}
so
\begin{equation}
V=\frac{q}{4\pi\epsilon_0}\frac{(s^2+z^2+b^2)}{z\sqrt{(s^2+z^2+b^2)^2-(2bz)^2}}\theta(z),
\end{equation}
and we recover Eq.~24 (for $t=0$).  Still no sign of the extra term in Eq.~32; evidently the retarded potentials {\it themselves} are incorrect, in this case.  

\section{What Went Wrong?}
Straightforward application of the standard formulas for the field (Eq.~2), the Li\'enard-Weichert potentials (Eq.~22), and the retarded potential (Eq.~34), yield incorrect results (inconsistent with Maxwell's equations) in the case of a charged particle in hyperbolic motion---they all miss an essential delta-function contribution.  How did this happen?  Bondi and Gold\footnote{Reference 7, quoted in Eriksen and Gr{\o}n, ref.~2.} write,
\begin{quotation} ``The failure of the method of retarded potentials to give the correct field is hardly surprising.  The solution of the wave equation by retarded potentials is valid only if the contributions due to distant regions fall off sufficiently rapidly with distance."
\end{quotation}
Fulton and Rohrlich\footnote{Reference 12, quoted in Eriksen and Gr{\o}n, ref.~2.} write,
\begin{quotation}
``The Li\'enard/Wiechert potentials are not valid in the present case at $T+z=0$, because their derivation assumes that the source is {\it not} at infinity."
\end{quotation}
But where, exactly, do the standard derivations make these assumptions, and how can they be generalized to cover the hyperbolic case?\footnote{Lyle (ref.~2, page 216) thinks it ``likely" that the extra terms could in fact be obtained at the level of the Li\'enard-Wiechert potentials ``if we were more careful about the step function," but he offers no justification for this conjecture.}  Zangwill\footnote{A.~Zangwill, {\it Modern Electrodynamics}, Cambridge University Press, Cambridge (2013), Section 20.3.} offers a careful, step-by-step derivation of the retarded potentials; one of those steps must fail, but  we have been unsuccessful in identifying the guilty party.  And although it is easy to construct configurations for which the retarded potentials break down, we know of no other case for which the field formula (Eq.~2) fails.

\bigskip

\noindent{\Large{\bf Acknowledgement}}
We thank Colin LaMont for introducing us to this problem.\footnote{C.~LaMont, ``Relativistic Direct Interaction Electrodynamics: Theory and Computation," Reed College senior thesis, 2011.}

\vfill\eject

\centerline{\bf {\LARGE Appendices}}

\appendix

\section{Retarded time for points on the $xy$ plane.}

The retarded time for a point in the $xy$ plane, at time $t$, is given by
\begin{equation}
c(t-t_r) = \sqrt{x^2+y^2+z(t_r)^2} = \sqrt{x^2+y^2+b^2+c^2t_r^2}\, ,
\end{equation}
or
$$c^2t^2-2c^2tt_r+c^2t_r^2=x^2+y^2+b^2+c^2t_r^2,$$
so
\begin{equation}
t_r=\frac{c^2t^2-x^2-y^2-b^2}{2c^2t} = \frac{t^2-a^2}{2t},\quad{\rm where}\quad a^2\equiv \frac{x^2+y^2+b^2}{c^2}.
\end{equation}

\vskip0in
\begin{figure}[t]
\hskip-.8in\scalebox{1.0}[1.0]{\includegraphics{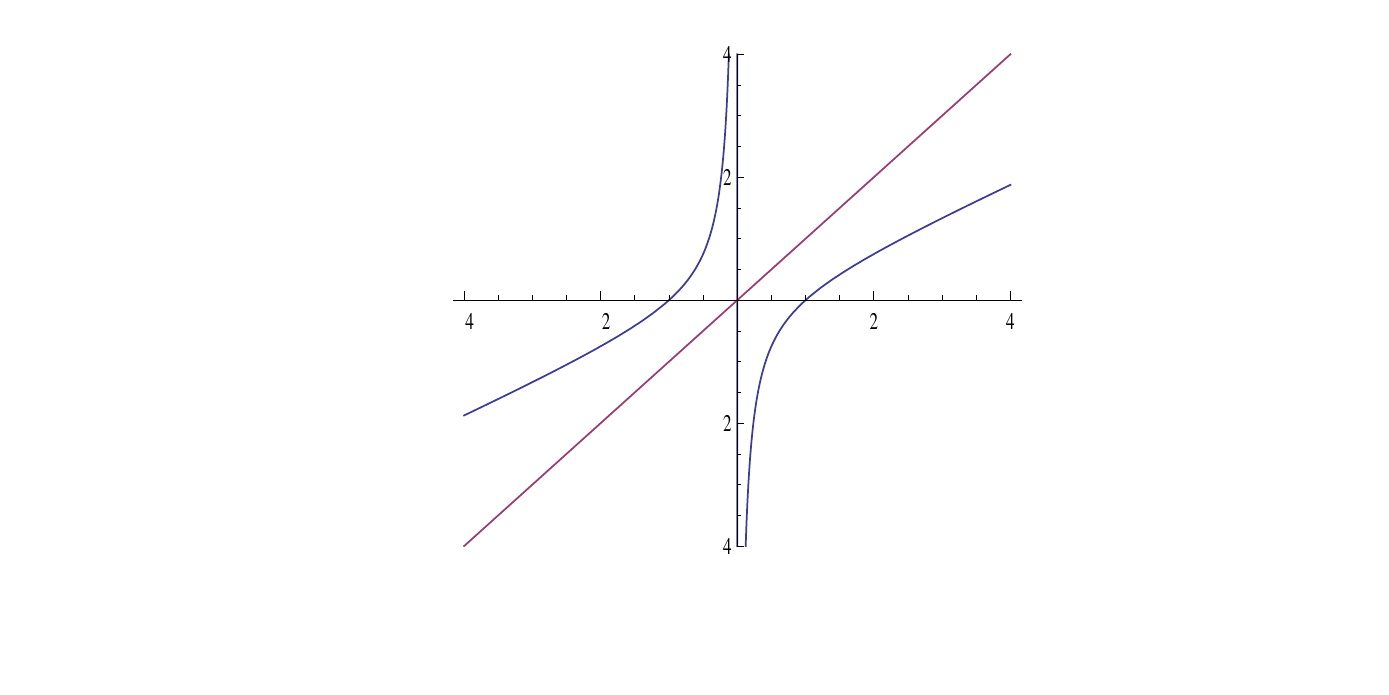}}
\vskip-.5in
\caption{Graph of the retarded time, as a function of $t$, for points in the $xy$ plane.}
\end{figure}

\noindent In Figure 6, $t_r$ is plotted (as a function of $t$), for $a=1$.  It is clear that $t_r<t$ for all positive $t$, but $t_r>t$ for all negative $t$.  The latter is no good, of course, but we do get an acceptable solution for all $t\geq 0$.  For $t=0$ the retarded time is (minus) infinity, regardless of the values of $x$ and $y$.

\section{Potentials.}
The vector from the (retarded) position of the charge to the point ${\bf r}=({\bf s},z)$, is
\begin{equation}
\brcurs = {\bf s} +\left(z-\sqrt{b^2+T_r^2}\right)\,\khat.
\end{equation}
The retarded time is given by
\begin{equation}
T-T_r = \rcurs = \sqrt{s^2+\left(z-\sqrt{b^2+T_r^2}\right)^2}.
\end{equation}
Squaring twice and solving the resulting quadratic yields Eq.~23.\footnote{The sign of the radical is enforced by the condition $T>T_r$.}

Referring back to Eqs.~3 and 5, the denominator in Eq.~22 is 
\begin{equation}
d \equiv \rcurs - \frac{\brcurs\cdot {\bf v}}{c} = (T-T_r)  - \left(z-\sqrt{b^2+T_r^2}\right)\frac{T_r}{\sqrt{b^2+T_r^2}}=T-\frac{zT_r}{\sqrt{b^2+T_r^2}}.
\end{equation}
It pays to use Eq.~49 to eliminate the radical:
\begin{equation}
2z\sqrt{b^2+T_r^2}=(s^2+z^2+b^2-T^2) + 2TT_r,
\end{equation}
so
\begin{equation}
d = T - \frac{2z^2T_r}{(s^2+z^2+b^2-T^2) + 2TT_r}.
\end{equation}
Putting in Eq.~23, and simplifying, 
\begin{equation}
d=\frac{(T^2-z^2)\sqrt{(s^2+z^2+b^2-T^2)^2+4b^2(T^2-z^2)}}{T\sqrt{(s^2+z^2+b^2-T^2)^2+4b^2(T^2-z^2)}-z(s^2+z^2+b^2-T^2)},
\end{equation}
so
$$V= \frac{q}{4\pi\epsilon_0}\frac{1}{d}  = \frac{q}{4\pi\epsilon_0}\frac{1}{(T^2-z^2)}\left[T -\frac{z(s^2+z^2+b^2-T^2)}{\sqrt{4b^2(T^2-z^2)+(s^2+z^2+b^2-T^2)^2}}\right]$$
(Eq.~24).

Meanwhile, the vector potential (Eq.~22) is
\begin{eqnarray*}
{\bf A}& =& \frac{{\bf v}}{c^2}V = \frac{T_r}{c\sqrt{b^2+T_r^2}}V\,\khat= \frac{2zT_r}{c[(s^2+z^2+b^2-T^2) + 2 T T_r]}\left(\frac{q}{4\pi\epsilon_0 d}\right)\khat\\
&=&\left(\frac{q}{4\pi\epsilon_0 c}\right)\frac{2zT_r}{T[(s^2+z^2+b^2-T^2)+2TT_r]-2z^2T_r}\,\khat\\
&=&\frac{q}{4\pi\epsilon_0 c}\frac{1}{(T^2-z^2)}\left[z-T\frac{(s^2+z^2+b^2-T^2)}{\sqrt{4b^2(T^2-z^2)+(s^2+z^2+b^2-T^2)^2}}\right]\,\khat
\end{eqnarray*}
(Eq.~25).

\section{Radiation}
From the potentials (Eqs.~32 and 33) we obtain the fields:\footnote{Any reader with lingering doubts is invited to check that these fields satisfy all of Maxwell's equations.  Note the critical role of the delta functions in Gauss's law and the Amp\`ere-Maxwell law.}
\begin{eqnarray}
{\bf E}(s,\phi,z,t) &=&\frac{qb^2}{\pi\epsilon_0}\left\{\frac{\left(z^2-s^2-b^2-T^2\right)\,\khat + 2z\,{\bf s}}{\left[(z^2+s^2+b^2-T^2)^2-4b^2(z^2-T^2)\right]^{3/2}}\right\}\theta(z+T)\nonumber\\
&& +\,  \frac{q}{2\pi\epsilon_0}\frac{{\bf s}}{s^2+b^2}\delta(z+T);\\
{\bf B}(s,\phi,z,t) &=&\Bigg\{\frac{qb^2}{\pi\epsilon_0c}\frac{2Ts}{\left[(z^2+s^2+b^2-T^2)^2-4b^2(z^2-T^2)\right]^{3/2}}\theta(z+T)\nonumber\\
&& -\,  \frac{q}{2\pi\epsilon_0c}\frac{s}{s^2+b^2}\delta(z+T)\Bigg\}\phat = \frac{1}{c}(\hrcurs \times {\bf E}).
\end{eqnarray}

To calculate the power radiated by the charge at a time $t_r$ (when it is located at the point $z(t_r)=\sqrt{b^2-T_r^2}$), we integrate the Poynting vector,
\begin{equation}
{\bf S} = \frac{1}{\mu_0}({\bf E} \times{\bf B}),
\end{equation} 
over a sphere of radius $\rcurs = T-T_r$ centered at $z(t_r)$, and take the limit as $\rcurs\to\infty$, with $T_r=ct_r$ held constant.  (That is, we track the energy as it flows outward at the speed of light; ``radiation" is the portion that makes it ``all the way to infinity.")  We need the fields, then, at later and later times, as the sphere expands.  Now, the delta-function term is confined to the plane $z=-T$, which recedes farther and farther to the left, as time goes on (Figure 7), and the expanding sphere never catches up.  Curiously, then, the delta-function term does not contribute to the power radiated by the charge at any (finite) point on its trajectory.  By the same token, the spherical surface is always in the region where $z+T>0$, so we can drop the theta functions.

Now,
\begin{equation}
{\bf S} = \frac{1}{\mu_0}({\bf E} \times{\bf B}) = \frac{1}{\mu_0c}[{\bf E} \times(\hrcurs \times{\bf E})] = \frac{1}{\mu_0c}[E^2\hrcurs - (\hrcurs \cdot {\bf E}){\bf E}],
\end{equation}
\noindent and on the surface of the sphere $d{\bf a} = \rcurs^2\,\sin\theta\,d\theta\,d\phi\,\hrcurs$:
\begin{equation}
{\bf S}\cdot d{\bf a} = \frac{1}{\mu_0c}[E^2\rcurs^2 - (\brcurs \cdot {\bf E})^2] \,\sin\theta\,d\theta\,d\phi.
\end{equation}
The power radiated is\footnote{The factor $(1-\brcurs\cdot {\bf v}/\rcurs c)$ accounts for the fact that the rate at which energy leaves a (moving) charge is not the same as the rate at which it (later) crosses a patch of area on the sphere.  See ref.~3, page 485.}
\begin{equation}
P = \lim_{\rcurs \to \infty}\left\{\frac{1}{\mu_0c}\int\left(1-\frac{\brcurs\cdot{\bf v}}{\rcurs c}\right) [E^2\rcurs^2 - (\brcurs \cdot {\bf E})^2] \,\sin\theta\,d\theta\,d\phi\right\}.
\end{equation}
Using the relevant fields (Eq.~54) we find
\begin{equation}
P = \lim_{\rcurs \to \infty}\left\{\left(\frac{\rcurs + T_r}{\rcurs}\right)^2\frac{cq^2}{6\pi \epsilon_0b^2}\right\}
= \frac{cq^2}{6\pi\epsilon_0 b^2}.
\end{equation}
Perhaps surprisingly, it is {\it constant} (independent of $T_r$, and hence the same for all points on the trajectory).\footnote{The fact that a charged particle in hyperbolic motion radiates has interesting implications for the equivalence principle---in fact, it is this aspect of the problem that has attracted the attention of most of the authors cited here.  Incidentally, the particle experiences {\it no} radiation reaction force---see R.~Peierls, {\it Surprises in Theoretical Physics} (Princeton University Press, Princeton, NJ, 1979, Chapter 8.)}  It agrees with the Li\'enard formula\footnote{Reference 3, Eq.~11.73.} (for collinear {\bf v} and {\bf a}),
\begin{equation}
P = \frac{q^2}{6\pi \epsilon_0 c^3}\gamma^6a^2.
\end{equation}

\vfill\eject

\vskip0in
\hskip0in\scalebox{.7}[.7]{\includegraphics{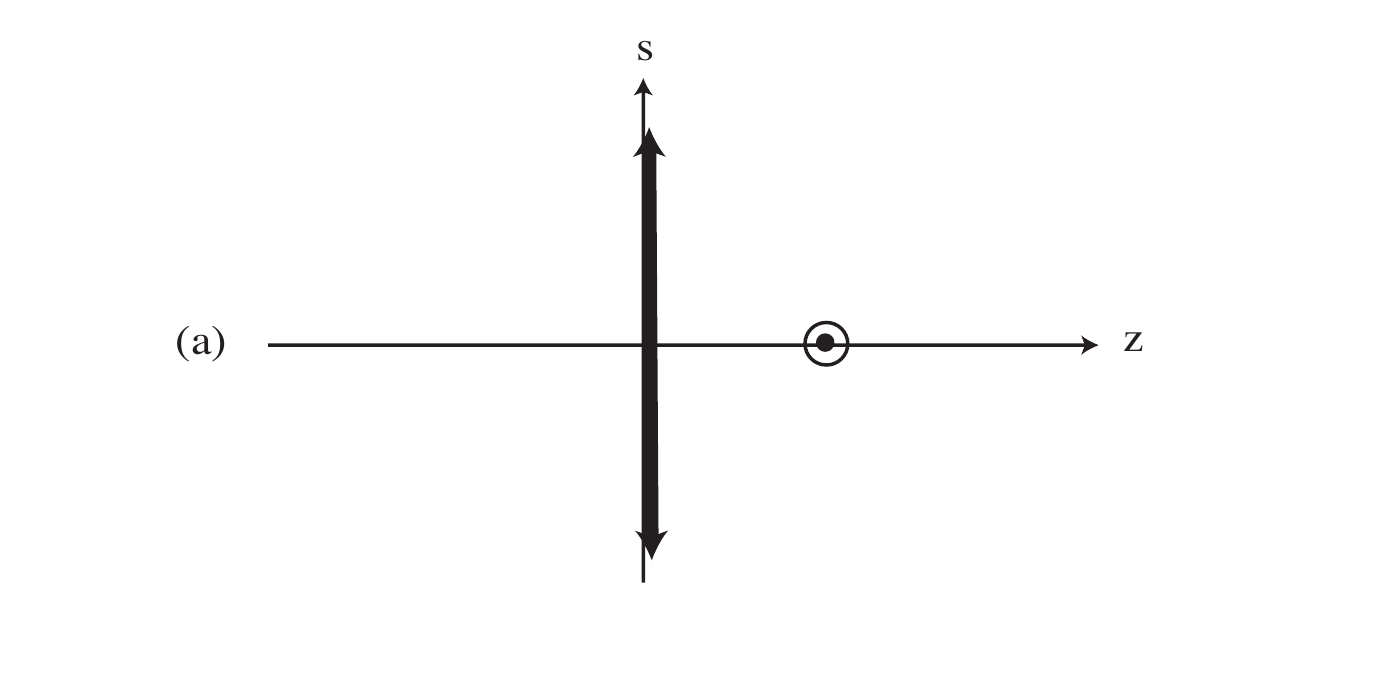}}
\vskip-.2in

\hskip0in\scalebox{.7}[.7]{\includegraphics{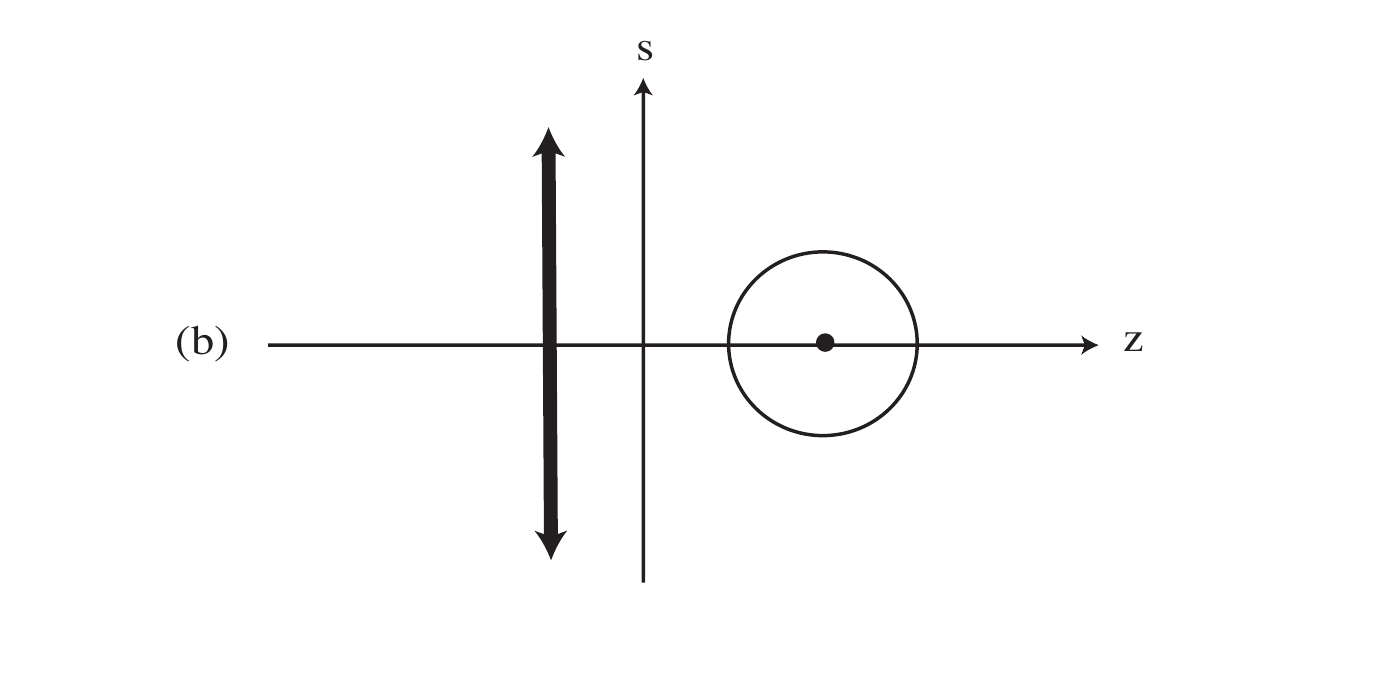}}
\vskip-.3in

\begin{figure}[h]
\hskip.2in\scalebox{.7}[.7]{\includegraphics{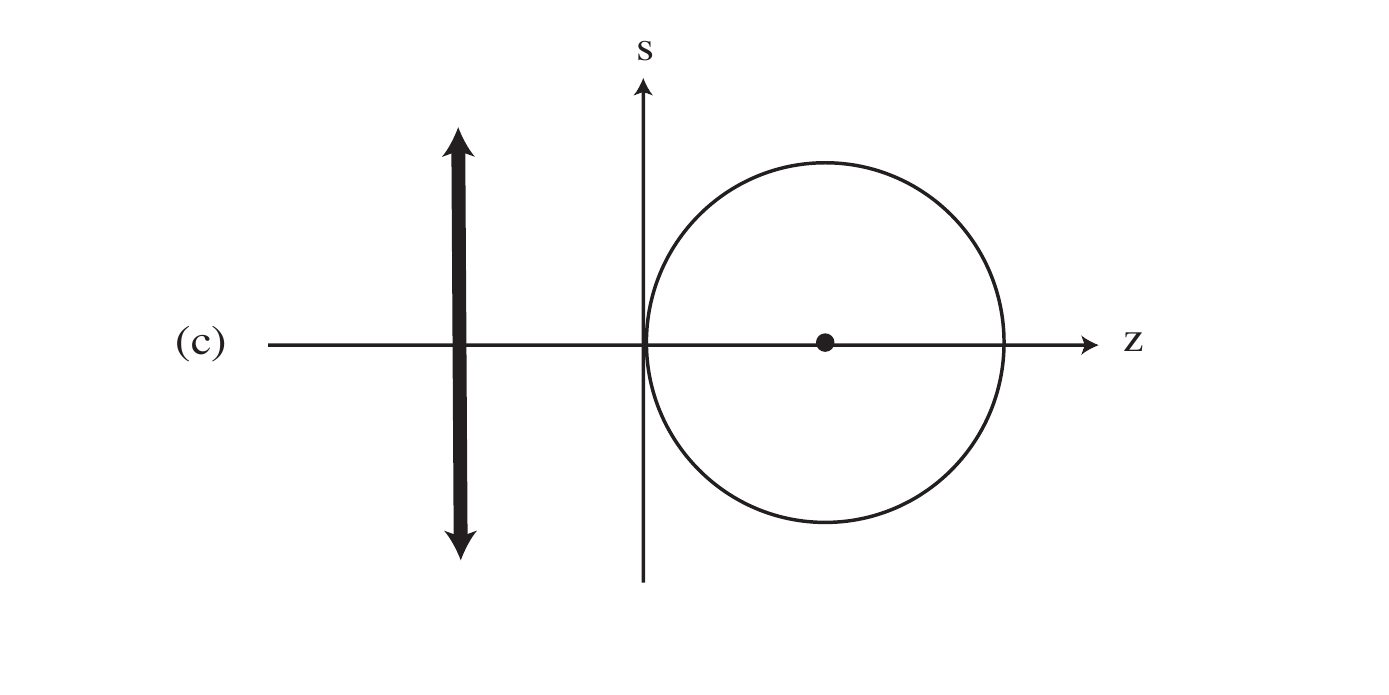}}
\vskip-.2in
\caption{Radiation from the charge at time zero (for $b=1$), showing the spherical surface and the delta-fields at (a) $T =0$, (b) $T  = 1/2$, (c) $T=1$.}
\end{figure}

\end{document}